\documentclass[12pt]{article}
\usepackage[utf8]{inputenc}
\usepackage{amsmath,amssymb}
\usepackage{geometry}
\usepackage{hyperref}
\usepackage{cite}
\usepackage{xcolor}
\definecolor{sehrdunkelblau}{rgb}{0.0, 0.0, 0.3} 

\begin{document}	
	\title{The Extended Uncertainty Principle from a Projector-Valued Measurement Perspective}
	\author{Thomas Sch\"urmann\\D\"usseldorf, Germany}
	\date{}
	\maketitle
	
	\begin{abstract}
	We revisit the Extended Uncertainty Principle (EUP) from an operational viewpoint, replacing wavefunction-based widths with apparatus-defined position constraints such as a finite slit of width $\Delta x$ or a geodesic ball of radius $R$. Using Hermitian momentum operators consistent with the EUP algebra, we prove a sharp lower bound on the product of momentum spread and preparation size in one dimension and show that it reduces smoothly to the standard quantum limit as the deformation vanishes. We then extend the construction to the dimensions two and three on spaces of constant curvature and obtain the corresponding bound for spherical confinement, clarifying its geometric meaning via an {\it isometry} to $S^2$ and $S^3$. The framework links curvature-scale effects to operational momentum floors and suggests concrete tests in diffraction, cold-atom, and optomechanical settings.
	\end{abstract}	

	\vspace{2ex}
	
	\vspace{2ex}

	\section{Introduction}
	
	In recent decades, the search for a consistent theory of quantum gravity has led to various modifications of the Heisenberg Uncertainty Principle (HUP). Among the most widely studied approaches are the Generalised Uncertainty Principle (GUP) and the Extended Uncertainty Principle (EUP), both of which introduce corrections to the standard quantum mechanical uncertainty relation to account for quantum gravitational effects \cite{Kempf1995}\cite{Das2008}\cite{Scardigli1999}\cite{ACBMP24}. These frameworks often incorporate minimal length scales or other geometrical considerations that occur naturally in many theories of gravity at very small scales, such as string theory and loop quantum gravity \cite{Garay1995}\cite{Rovelli1998}. 	
	The GUP generally modifies the canonical commutation relation between position and momentum by adding terms that become significant at very high energies or near the Planck scale \cite{Adler2001}\cite{Das2009}. An important result of this modification is a fundamental limit to the spatial resolution, sometimes referred to as the "minimum length", which prevents measurements below a certain scale \cite{Kempf1995}\cite{Scardigli1999}. Such a minimal length concept is seen as a natural consequence of the interplay between quantum mechanics and gravity, suggesting that classical notions of spacetime break down at the Planck scale \cite{Garay1995}.
	
	In contrast, the EUP arises in scenarios where curvature effects or large-scale structures become relevant, typically in cosmological or de Sitter-like spacetimes \cite{Park2007}. Rather than focusing on short-range modifications, the EUP addresses long-range corrections to the HUP that may arise in an expanding universe or in contexts where the cosmological constant and horizon scales cannot be neglected \cite{Bolen2005}\cite{Mignemi2010}\cite{KSH2024}. These corrections indicate a limit to how accurately one can measure the momentum when the expansion of the universe is taken into account.
	In this picture, a position-dependent term is added to the canonical commutator
	\begin{equation}
		[\hat x, \hat p] = i\hbar \bigl( 1 + \alpha\, \hat x^2 \bigr),
		\label{mcom}
	\end{equation}
	where $\alpha$ is a parameter typically associated with cosmological effects
	 - for instance, by taking $\alpha = 1/L_H^{2}$ with $L_H = c/H_0$ the Hubble radius and $H_0$ the Hubble constant, i.e., the present‑day value of the Hubble parameter.

	From a thermodynamic perspective, enlarging the positional cell in phase space and introducing a ``minimum momentum'' reduces the available phase-space volume for any finite macroscopic system. This in turn decreases entropy and related thermodynamic potentials, and --- when combined with a minimal-length GUP --- can render both the ultraviolet (UV) and infrared (IR) sectors finite \cite{MAZ2021}\cite{Pachol2025}. This physical and thermodynamic picture is reflected directly in the mathematical structure of the commutator.

	Proceeding much as in the GUP case, one finds
	\begin{equation}
		\sigma_p\,\sigma_x \ge \frac{\hbar}{2}
		\Bigl[\,1 + \alpha \, \sigma_x^{2}\Bigr].
		\label{eup}
	\end{equation}
	Here, large values of $\sigma_x$ induce noticeable deviations from the usual uncertainty relation, suggesting that additional quantum effects may be at play at very large (cosmological) distances.
	To derive the standard EUP inequality from the modified commutation relation, one typically follows the proof of the usual uncertainty relation by applying the Cauchy-Schwarz inequality or the Robertson-Schr\"odinger approach. For any two arbitrary (self-adjoint) operators $\hat{A}$ and $\hat{B}$,
	\begin{equation}
		\sigma_{A}\,\sigma_{B} \ge \frac{1}{2}\,\bigl|\langle [\hat{A},\,\hat{B}]\rangle \bigr|,
	\end{equation}
	where $\sigma_{A}$ and $\sigma_{B}$ are the standard deviations of $\hat{A}$ and $\hat{B}$ in the state of interest. Applying this to $\hat{x}$ and $\hat{p}$, with their modified commutators, we obtain a relation that contains additional terms from the factor $\bigl(1 + \alpha\,\hat{x}^2\bigr)$.\\

	It is worth noting, however, that this derivation is not fully rigorous from a mathematical standpoint. In particular, it often relies on the simplifying assumption $\langle \hat x \rangle = 0$ and does not completely account for the operator nature of  $(\hat x-\langle \hat x\rangle)^2$. A direct application of the Robertson–Schrödinger inequality actually yields 
	\begin{equation}
		\sigma_p \,\sigma_x
		\geq
		\frac{\hbar}{2}\,\Bigl[\,1 + \alpha\,(\sigma_x^2+\langle\hat x\rangle^2)\Bigr],
	\end{equation}
	which differs from (\ref{eup}) by an additional, state-dependent term in the bracket. Since many derivations in the literature assume $\langle \hat{x} \rangle = 0$ and simplify the operator structure of $(\hat{x} - \langle \hat{x}\rangle)^2$, the resulting conclusions should be regarded as suggestive rather than fully rigorous.
	
	In order to obtain an inequality whose right-hand side is strictly independent of the spatial origin, one must reconsider the suitability of the classical uncertainty measure $\sigma_x$. Unlike the standard deviation~$\sigma_x$, which can be strongly influenced by the shape and tails of the wavefunction, the slit width~$\Delta x$ provides a direct indicator of the spatial constraint imposed by the experimental setup. 		
	From the perspective of projector-valued measurements, $\Delta x$ identifies the actual range over which the position projector acts, thereby capturing the physical confinement imposed by the apparatus. In this operational picture, $\sigma_p\Delta x$ becomes a directly measurable and conceptually transparent uncertainty product. As shown in earlier work \cite{TS09,TS23}, imposing projector-valued preparation on wavefunctions with suitable boundary conditions leads to a fundamental lower bound
	\begin{equation}
	\sigma_p\Delta x \geq \pi\,\hbar, 
	\end{equation}
	illustrating that $\Delta x$ can serve as the primary position measure in an uncertainty relation that is both rigorous and experimentally meaningful.\\

	This paper is organised as follows. In Section\,\ref{proof}, we consider the one-dimensional EUP momentum operator and derive a strict lower bound on the product $\sigma_p \Delta x$ (Theorem~1). Next, in Section\,\ref{proof2} we extend the analysis to a finite spherical region of radius $R$, representing the position uncertainty, and establish an analogous bound in three dimensions (Theorem~2). 
	Section\,\ref{gen} generalises the approach to arbitrary spatial dimension $d$, including an explicit treatment of the two-dimensional, non-parallelizable case on $S^2$. Section\,\ref{con} provides concluding remarks and an outlook.
			
	\section{The 1D EUP: A Strict Lower Bound on $\sigma_p \Delta x$}\label{proof}
	\noindent
	
	The following theorem applies the Hermitian EUP momentum operator discussed in \cite{IM20}. By imposing boundary conditions that mimic a projector-valued preparation in $\Delta x$, we demonstrate the emergence of a new, fundamental limit:\\
	
	\noindent
	\textbf{Theorem 1.} 
	Let $\Delta x > 0$ and $\alpha \geq 0$ be real parameters. Consider the Hermitian EUP momentum operator
	\begin{equation}
		\hat{p} = -\,i\hbar\,\bigl[(1 + \alpha\,x^2)\,\tfrac{d}{dx} + \alpha\,x\bigr].
		\label{eup_op}
	\end{equation}
	Then the following fundamental lower bound applies:
	\begin{equation}\label{un1d}
		\sigma_p\,\Delta x \geq \pi\hbar\;\Phi_\alpha(\tfrac{\Delta x}{2}),
	\end{equation}
	where
	\begin{equation}\label{Phi}
		\Phi_\alpha(z)
		=
		\frac{\sqrt{\alpha}\,z}{\arctan\bigl(\sqrt{\alpha}\,z\bigr)}.
	\end{equation}
	\\
	\textbf{Remark.} 
	\\
	\begin{equation}\label{appr_un}
		\sigma_p\,\Delta x \geq\pi\hbar\; \Bigl[\,1+\frac{\alpha\, \Delta x^2}{12}+{\cal O}\bigl(\alpha^2\Delta x^4\bigr)\Bigr].
	\end{equation}
	\\
	\noindent
	\textbf{Proof.} In order to determine the underlying Hilbert space, we begin with a brief examination of the eigenvalue equation:
	\begin{equation}\label{ewg}
		\hat{p}\,\psi(x) = p\,\psi(x),
	\end{equation}
	where $p \in \mathbb{R}$ is the real eigenvalue identified with the physical momentum. 
	By applying the definition in \eqref{eup_op} to \eqref{ewg}, we obtain the differential equation
	\begin{equation}
		-\,i\hbar\,\bigl[(1 + \alpha x^2)\,\psi'(x) + \alpha\,x\,\psi(x)\bigr] 
		= 
		p\,\psi(x).
		\label{eup_dgl}
	\end{equation}
	Dividing both sides by $-\,i\hbar$ and by $(1 + \alpha x^2)\,\psi(x)$ leads to the condition:
	\begin{equation}
		\frac{\psi'(x)}{\psi(x)}  
		= 
		\frac{\tfrac{p}{i\hbar} - \alpha x}{1 + \alpha x^2}.
		\label{eup_log}
	\end{equation}
	This form can be integrated directly using the logarithmic derivative approach, referring to a result often found in the literature which shows that the eigenfunctions have the following form
	\begin{equation}
		\psi_p(x)
		=
		C \,
		\frac{1}{\sqrt{1 + \alpha x^2}}\,
		\exp\Bigl(
		-\,i\,\frac{p}{\hbar\,\sqrt{\alpha}}\;\arctan(\sqrt{\alpha}\,x)
		\Bigr).
	\end{equation}
	As $|x|\to\infty$, we have $\arctan(\sqrt{\alpha} \, x) \to \pm \tfrac{\pi}{2}$ and $(1 + \alpha x^2)^{-1/2} \sim 1/(\sqrt{\alpha}\,|x|)$. Therefore,
	\begin{equation}
		\bigl|\psi_p(x)\bigr|^2
		\sim
		\frac{1}{\alpha\,x^2},
	\end{equation}
	which can be integrated over the real line. 
	Thus these wavefunctions are square integrable (normalisable) for $\alpha>0$. 
	Despite the deformed commutation relation, $p \in \mathbb{R}$ remains a continuous parameter, analogous to the standard momentum continuum in ordinary quantum mechanics.  Unlike the usual plane waves $e^{i p x / \hbar}$, these eigenfunctions prove to be normalisable thanks to the factor $(1+\alpha x^2)^{-1/2}$.
	In the limit $\alpha \to 0$ we have $\arctan(\sqrt{\alpha}\,x) \sim \sqrt{\alpha}\,x$ for moderate $|x|$, and 
	\begin{equation}
		\psi_p(x)
		\longrightarrow
		\exp\Bigl(-\,i\,\tfrac{p}{\hbar}\,x\Bigr),
	\end{equation}
	which recovers the standard plane-wave solutions (up to normalisation factors) of the usual momentum operator $\hat{p}=-\,i\hbar\,\tfrac{d}{dx}$. This is consistent with the standard commutation relation $[x,p]=i\hbar$, which reappears as $\alpha\to 0$.\\
	
	Now, for a real parameter $p$, we introduce two types of functions, $\Psi_{p}^+(x)$ and $\Psi_{p}^-(x)$, which are derived from the same set of base functions $\psi_{p}(x)$. The plus sign leads to a symmetric combination of the functions $\psi_{p}(x)$ and $\psi_{-p}(x)$, while the minus sign leads to an anti-symmetric combination. Specifically, the symmetric function is given by
	\begin{equation}
		\Psi_{p}^+(x) = \psi_{p}(x) + \psi_{-p}(x),
	\end{equation}
	while the antisymmetric function is set to
	\begin{equation}
		\Psi_{p}^-(x) = \psi_{p}(x) - \psi_{-p}(x).
	\end{equation}
	These definitions ensure that $\Psi_{p}^+(x)$ remains unchanged under the transformation $x \mapsto -x$, while $\Psi_{p}^-(x)$ changes its sign under the same transformation, highlighting their symmetric and antisymmetric properties respectively.
	Then $\Psi_p^\pm$ is an eigenfunction of $\hat{p}^2$ with the eigenvalue $p^2$ according to 
	\begin{equation}
		\hat{p}^2\,\Psi_p^\pm(x) = p^2\,\Psi_p^\pm(x).
	\end{equation}
	Note that neither $\Psi^+_{p}(x)$ nor $\Psi^-_{p}(x)$ can be an eigenfunction of both $\hat{p}$ and $\hat{p}^2$ with the same eigenvalue. However, $\psi_{p}(x)$ and $\psi_{-p}(x)$ have the same eigenvalue $p^2$ for the operator $\hat{p}^2$.
	Consequently, any linear combination $\psi_p(x) + \psi_{-p}(x)$ (or a more general combination $a\,\psi_p + b\,\psi_{-p}$) is still an eigenstate of $\hat{p}^2$. This parallels the standard quantum mechanical fact that plane waves $e^{i k x}$ for $+k$ and $-k$ have the same value of $\hat{p}^2$, corresponding to the same $p^2 = (\hbar k)^2$ in the free particle case.\\
	
	\noindent
	Now we impose the Dirichlet boundary conditions 
	\begin{equation}
		\Psi_{p_n}^+\Bigl(\pm\tfrac{\Delta x}{2}\Bigr) = 0 
		\quad\text{and}\quad 
		\Psi_{p_n}^-\Bigl(\pm\tfrac{\Delta x}{2}\Bigr) = 0\,.
	\end{equation}
	This condition ensures that the standard deviation of the momentum, $\sigma_p$, remains finite. We find that the discrete set of solutions then has the form
	\begin{equation}
		\label{wave}
		\Psi_{p_n}^{\pm}(x) 
		= 
		\frac{C}{\sqrt{\,1 + \alpha\,x^{2}\,}}
		\begin{cases}
			\displaystyle
			\cos\Bigl(\frac{n\pi}{2}\,\frac{\arctan(\sqrt{\alpha}\,x)}{\arctan(\sqrt{\alpha}\,\Delta x/2)}\Bigr), 
			& \text{for odd }n,\\[1em]
			\displaystyle
			\sin\Bigl(\frac{n\pi}{2}\,\frac{\arctan(\sqrt{\alpha}\,x)}{\arctan(\sqrt{\alpha}\,\Delta x/2)}\Bigr), 
			& \text{for even }n,
		\end{cases}
		\quad n \neq 0,
	\end{equation}
	over the finite interval $\bigl[-\tfrac{\Delta x}{2},\,\tfrac{\Delta x}{2}\bigr]$. Note that the argument of the trigonometric functions is rescaled so that the boundary conditions map exactly to integer multiples of $\pi$. The choice of 
	$\cos$ or $\sin$ depends on the parity of $n$. \\
	\\
	The corresponding point spectrum of the momentum operator is thus given by
	\begin{equation}
		p_n 
		=
		\frac{\pi\hbar \,\sqrt{\alpha}/2}{\arctan\bigl(\sqrt{\alpha}\,\tfrac{\Delta x}{2}\bigr)} 
		\times n,
		\quad\quad
		n\in \mathbb{Z}\setminus \{0\}.
	\end{equation}
	The factor $n\pi/2$ in the trigonometric function in (\ref{wave}) ensures that the boundary values at $\pm \Delta x/2$ vanish, giving a discrete set of allowed momenta $\{p_n\}$. For 
	\begin{equation}
		C = \left[
		\frac{\sqrt{\alpha}}{\arctan\bigl(\sqrt{\alpha}\,\frac{\Delta x}{2}\bigr)}
		\right]^{\frac{1}{2}},
	\end{equation}
	the eigenfunctions $\Psi_{p_n}^\pm(x)$ form an orthonormal system on $L^2\bigl([-\tfrac{\Delta x}{2},\tfrac{\Delta x}{2}]\bigr)$:
	\begin{equation}
		\langle \Psi_{n}^\pm, \Psi_{m}^\pm \rangle = \delta_{nm},
		\quad
		\langle \Psi_{n}^\pm, \Psi_{m}^\mp \rangle = 0
		\quad
		\text{for all } n,m \in \mathbb{Z} \setminus \{0\}.
	\end{equation}
	A convenient strategy for the proof of completeness is the transformation of the problem into a standard trigonometric system on a \emph{dimensionless} interval. Let us define
	\begin{equation}
		U = \arctan\bigl(\sqrt{\alpha}\,\tfrac{\Delta x}{2}\bigr),
	\end{equation}
	and introduce
	\begin{equation}
		u = \frac{\arctan\bigl(\sqrt{\alpha}\,x\bigr)}{U}.
	\end{equation}
	So, just as $x$ goes from $-\Delta x/2$ to $+\Delta x/2$, the variable $u$ goes from $-1$ to $+1$. Moreover,
	\begin{equation}
		\frac{1}{\sqrt{1 + \alpha x^2}}=\cos\bigl(U u\bigr),
		\quad
		\arctan\bigl(\sqrt{\alpha}\,x\bigr) = U u.
	\end{equation}
	Substituting $\arctan(\sqrt{\alpha}\,x) = U\,u$ into \eqref{wave} leads directly to
	\begin{equation}
		\Psi_{p_n}^\pm\bigl(x(u)\bigr)= C \,\cos \bigl(U u\bigr)
		\begin{cases}
			\cos\bigl(\tfrac{n\pi}{2}\,u\bigr), & n \text{ odd},\\[6pt]
			\sin\bigl(\tfrac{n\pi}{2}\,u\bigr), & n \text{ even}.
		\end{cases}
	\end{equation}
	The map $x\mapsto u$ is a smooth bijection from $\bigl[-\tfrac{\Delta x}{2},\,\tfrac{\Delta x}{2}\bigr]$ to $[-1,1]$. Since sines and cosines form a complete basis in $L^2([-1,1])$, their completeness is lifted back to the original $x$ interval under this bijection. Consequently, the wave functions $\{\Psi_{p_n}^\pm\}$ form a complete orthonormal system in $L^2\bigl[-\tfrac{\Delta x}{2},\,\tfrac{\Delta x}{2}\bigr]$.
	
	Finally, let us look at the smallest positive value of $p_n^2$. The mode with the smallest positive momentum, $p_n$, appears when $n = \pm 1$. Hence,
	\begin{equation}
		\min_n\, \lvert p_n\rvert
			=\frac{1}{2}\,\frac{\pi\,\hbar\,\sqrt{\alpha}}{\arctan\bigl(\sqrt{\alpha}\,\tfrac{\Delta x}{2}\bigr)}
		\,.
	\end{equation}
	All of the higher modes correspond to $\lvert n\rvert \ge 1$, which gives larger values of $\lvert p_n\rvert$ and thus larger $p_n^2$. The corresponding normalised wave function is
\begin{equation}
	\Psi^+_{p_1}(x) =\frac{\alpha^{1/4}}{\arctan\bigl(\sqrt{\alpha}\,\tfrac{\Delta x}{2}\bigr)^{1/2}}\,\,\,\frac{1}{\bigl(1+\alpha\, x^2\bigr)^{1/2}}\,
	\,\cos\Bigl(\frac{\pi}{2}\, \frac{\arctan(\sqrt{\alpha }\,x)}{\arctan(\sqrt{\alpha}\,\Delta x/2)}  \Bigr).
\end{equation}
Because of the symmetry, we find that
\begin{equation}
\langle \Psi^+_{p_1}\lvert\,\hat{p}\,\rvert\Psi^+_{p_1}\rangle = 0
\quad\text{and}\quad
\langle \Psi^+_{p_1}\lvert\,\hat{p}^2\,\rvert\Psi^+_{p_1}\rangle = p_1^2.
\end{equation}
It follows that $\sigma_p \,\ge\, |p_1|$, where equality holds for the state $\Psi^+_{p_1}(x)$. This completes the proof of our first theorem.
\hfill $\square$
\\
\\
The function $\Phi_\alpha(z)$
plays a crucial role in expressing the effect of large distance corrections on the position-momentum uncertainty. An intuitive interpretation of its behaviour is given below:	\\
	\\
\textbf{Small \boldmath$\alpha$ limit:} When we take the limit $\alpha \to 0$, our wavefunctions \eqref{wave} and their discrete eigenvalues naturally converge to the familiar quantum mechanical solutions - specifically plane waves in unbounded domains or discrete sines/cosines in finite intervals. In contrast, the approach in \cite{IM20} does not preserve this $\alpha \to 0$ limit.
Specifically, in Chapter III, Sec.\,A of \cite{IM20}, the derived energy eigenvalues $E_n(\alpha)$ and the corresponding wavefunctions $\phi_n(x;\alpha)$ both depend on the deformation parameter $\alpha$. According to this analysis, $\lim_{\alpha \to 0} E_n(\alpha) = 0$, $\lim_{\alpha \to 0} \phi_n(x;\alpha) = 0$.
Moreover, in \cite{IM20} the authors mention that $\hat{p}$ and $\hat{p}^2$ share the same eigenbasis. However, this is generally not the case and that $\hat{p}$ and $\hat{p}^2$ do not have a common eigenbasis.\\
\\		
\textbf{Large \boldmath$\alpha$ or \boldmath$\Delta x$ limit:} 
When $\sqrt{\alpha}\,\tfrac{\Delta x}{2}$ is large, $\arctan(\sqrt{\alpha}\,\tfrac{\Delta x}{2})$ approaches $\tfrac{\pi}{2}$. Consequently,
		\begin{equation}
			\Phi_\alpha(\tfrac{\Delta x}{2}) \;\sim\; 
			\frac{\sqrt{\alpha}\,\Delta x}{\pi},
		\end{equation}
suggesting that the correction factor $\Phi_\alpha(\Delta x)$ grows linearly with $\Delta x$. In this regime, gravitational (or cosmological) contributions become significant at large scales, leading to a minimum achievable standard deviation
		\begin{equation}
			\sigma_p \,\gtrsim \sqrt{\alpha}\,\hbar.
		\end{equation}
This result is consistent with the classical EUP approach, which uses $\sigma_x$.\\
\\
\textbf{Moderate \boldmath$\alpha$ values:} For intermediate regimes of $\alpha$ and $\Delta x$, $\Phi_\alpha(\tfrac{\Delta x}{2})$ deviates from unity but remains well below the linear growth implied by the large-$\alpha$ limit. This suggests a smooth interpolation between standard quantum behaviour and the fully extended scenario where cosmological effects strongly dominate.\\
\\
\textbf{Compatibility:} 
Although the wavefunction-based standard deviation $\sigma_x$ and the ap-paratus-based slit width $\Delta x$ are based on quite different foundations, there is a degree of compatibility between them. In particular, the strictly bounded $\sigma_x \le \Delta x/2$ underlines that although they are not mathematically identical concepts, they can be related in a consistent formal way. This does not establish a strictly rigorous equivalence - since $\sigma_x$ depends on the entire shape of the wavefunction in Hilbert space, whereas $\Delta x$ is purely geometric - but it does show that the two notions are not mutually exclusive. Indeed, one can construct a framework (e.g.\ $\Delta x = 2 \pi\sigma_x$) in which the EUP-type inequality (\ref{appr_un}) in $\Delta x$ is coherently translated into the inequality (\ref{eup}) in $\sigma_x$. This process involves the rescaling of the deformation parameter $\alpha\to 3\alpha/\pi^2$, which illustrates how $\Delta x$ and $\sigma_x$ remain compatible despite their conceptual differences.

\section{The 3D case EUP Lower Bound of $\sigma_p R$}\label{proof2}

Moving from one to three dimensions raises questions about rotational symmetry, tensor structures, and the choice of how the position operators and momenta should appear in the commutator. An obvious generalisation for the commutator in three dimensions is:
	\begin{equation}
		[\hat{x}_i, \hat{p}_j]
		=
		i\,\hbar\,\Bigl[\,
		\delta_{ij}\,(1
		+
		\alpha\,r^2)+\tilde\alpha\,\hat{x}_i\,\hat{x}_j 
		\Bigr],
	\end{equation}
where $i,j \in \{1,2,3\}$, $r^2 = \hat{x}_1^2 + \hat{x}_2^2 + \hat{x}_3^2$, and $\delta_{ij}$ is the Kronecker delta. This formulation respects isotropy in the sense that it captures both a purely radial term $r^2\,\delta_{ij}$ (isotropic part) and a directional contribution $\hat{x}_i\,\hat{x}_j$ (tensor part).\\
\\
An important consideration is how the modified commutator behaves under rotation. The inclusion of both $r^2\,\delta_{ij}$ and $\hat{x}_i\,\hat{x}_j$ ensures a framework consistent with spherical symmetry: $r^2\,\delta_{ij}$ remains rotationally invariant, while $\hat{x}_i\,\hat{x}_j$ captures directional anisotropies \cite{CH19} in a way that transforms appropriately under $SO(3)$. Although a formulation including both terms preserves rotational consistency and incorporates richer physics at large distances, it also adds complexity. In contrast, simpler radial-only versions may be adequate for certain applications, but may lose the full geometric richness of the three-dimensional setting. Ultimately, the choice depends on the physical context, the level of detail required and the feasibility of analytical or numerical approaches.

In the following, we focus mainly on analytical closed-form solutions, so we use the simpler extension according to
\begin{equation}
	[\hat{x}_i, \hat{p}_j]
	=
	i\,\hbar\,\delta_{ij}\, \bigl(1 + \alpha\,r^2\,\bigr),
	\label{eup_comm}
\end{equation}
without the term $\hat{x}_i\,\hat{x}_j$. 
In order to develop the corresponding quantum framework for the deformation parameter $\alpha$, a compatible three-dimensional Riemannian metric is constructed, and the associated Hermitian momentum operators in a different Hilbert space representation is applied. 
We start from the deformed classical Poisson bracket in coordinates $x_i$ and momenta $p_j$,
\begin{equation}
	\{x_i,p_j\} = \delta_{ij}\,(1+\alpha r^2),
\end{equation}
which corresponds to an inverse symplectic form with Liouville volume
\begin{equation}
	\mu_{\mathrm{Liouville}} = \frac{1}{(1+\alpha r^2)^3}\,d^3x\,d^3p.
\end{equation}
The spatial factor of this measure can be represented geometrically by the conformally flat metric
\begin{equation}
	g_{ij}(x) = \frac{\delta_{ij}}{(1+\alpha r^2)^2}, 
	\qquad
	\sqrt{g} = (1+\alpha r^2)^{-3}.
\end{equation}
Introducing the geodesic radial coordinate
\begin{equation}\label{physrad}
	r = \frac{1}{\sqrt{\alpha}}\tan(\sqrt{\alpha}\,\rho), 
	\qquad 0 \le \rho < \frac{\pi}{2\sqrt{\alpha}},
\end{equation}
the line element becomes
\begin{equation}
	ds^{2} = d\rho^{2} + \frac{1}{4\alpha}\sin^{2}\bigl(2\sqrt{\alpha}\,\rho\bigr)\,d\Omega^{2},
\end{equation}
with $d\Omega^2 = d\theta^2+\sin^2\!\theta\,d\varphi^2$.  
Defining
\begin{equation}
	a=\frac{1}{2\sqrt{\alpha}},
\end{equation}
yields
\begin{equation}\label{S3}
	ds^{2}= d\rho^{2} + a^{2}\sin^{2}\!\left(\frac{\rho}{a}\right)d\Omega^{2},
\end{equation}
the standard round metric on the 3-sphere $S^{3}$ of radius $a$. We denote it by $h = \Phi^{*}g$, with
\begin{equation}
	h_{ij}(x) = g_{\mu\nu}(\Phi(x))\,\partial_{i} \Phi^\mu\, \partial_{j} \Phi^\nu,
\end{equation}
where $g$ is the ambient metric and $\Phi$ the (isometric) embedding
\begin{equation}
	\Phi:\mathbb{R}^{3} \to S^{3}\setminus\{\mathrm{south\ pole}\}, \quad
	(r,\theta,\varphi) \mapsto (\rho,\theta,\varphi),
\end{equation}
with
\begin{equation}
	\rho = \frac{1}{\sqrt{\alpha}} \arctan(\sqrt{\alpha}\, r).
\end{equation}
The limit $\rho \to \pi a$ ($r\to\infty$) completes the embedding, compactifying $\mathbb{R}^3$ to $S^3$ with constant curvature
\begin{equation}
	K = \frac{1}{a^{2}} = 4\alpha.
\end{equation}
The Riemannian volume measure $d\mu=\sqrt{h}\,d\rho\, d\theta\, d\varphi$ matches the spatial factor of the Liouville measure.  
With the weighted inner product
\begin{equation}
	\langle \psi,\phi\rangle = \int \psi^*\phi\, d\mu,
\end{equation}
a Hermitian momentum operator must account for this nontrivial weight.
On the three-sphere $S^3$ one has the exceptional property of \emph{parallelizability}, 
which allows for the existence of a global, smooth, orthonormal frame of Killing vector fields 
$\{ X_i \}_{i=1}^3$ satisfying
\begin{equation}
	g(X_i,X_j) = \delta_{ij}, \quad \mathrm{div}\, X_i = 0 .
\end{equation}
This geometric fact has two important consequences for the quantum-mechanical definition of momentum variance. First, the momentum operators \cite{TS18}
\begin{equation}
	P_i = -i\hbar\, \nabla_{\!X_i}
\end{equation}
are automatically self-adjoint on $L^2(S^3,d\mu)$ without the need for additional 
connection terms of the De~Witt form \cite{dewitt52,dewitt57}.Here, $\nabla_{X_i}$ denotes the covariant derivative along $X_i$. Second, the Casimir operator
\begin{equation}
	P^2 = \sum_{i=1}^3 P_i^2 = -\hbar^2 \Delta_g
\end{equation}
is globally well-defined and coincides, up to the factor $-\hbar^2$, with the Laplace--Beltrami operator.
In this setting, the variance of the momentum can be written in the transparent component form
\begin{equation}\label{VarP}
	\sigma^2_p = 
	\sum_{i=1}^3 \big\langle \,(P_i  - \langle P_i \rangle)^2 \big\rangle
	= \langle P^2 \rangle - |\langle \mathbf{P} \rangle|^2 ,
\end{equation}
which is the direct analogue of the flat-space variance formula in $\mathbb{R}^3$.
Specifically, for a particle strictly localized in a geodesic ball of radius $R$ inside a simply connected 3-manifold of constant curvature $K=4\alpha$, there is the following statement:\\
\\
\textbf{Theorem 2.} \emph{The momentum standard deviation satisfies}
\begin{equation} \label{th2}
	\sigma_p R \geq \pi\hbar\,\left[1-\frac{4\alpha}{\pi^{2}}\, R^{2}\right]^\frac{1}{2},
\end{equation}
\emph{with $R\in [\,0,\pi/2\sqrt{\alpha}\,]$ for $\alpha>0$ (and $R\in [0,\infty)$ for $\alpha<0$) equality is attained by the first Dirichlet mode.}\\
\\
The proof is given in \cite{TS18}.
In the previously discussed case of constant positive curvature $K>0$, the spatial geometry was isometric to a 3-sphere $S^3$, leading to the compactness of space, a finite upper bound on the geodesic radius $R$, and a bound of the form (\ref{th2}).
However, for $K<0$ (constant negative curvature), the situation changes fundamentally: the spatial manifold is no longer compact but hyperbolic, with infinite spatial extent and no maximal geodesic radius. The inequality still has the same structural form (\ref{th2}), but now $R$ can take any positive value, and the curvature term inside the square root becomes $1 + |K| R^2/\pi^2$, leading to a bound that ``grows'' with $R$ instead of decreasing. 
Dividing by $R$ yields
\begin{equation} 
\sigma_p \;\xrightarrow{R\to\infty}\; \hbar\sqrt{|K|}\,.
\end{equation}
Thus, in hyperbolic space there is a curvature–set floor for the momentum uncertainty, i.e., a curvature–induced infrared momentum gap. This contrasts with the $K>0$ (spherical) case, where the bound on $\sigma_p$ can tend to zero as $R$ approaches its maximal geodesic radius.\\

In the 3D EUP framework, the parameter $R$ represents the ``geodesic radius'' on the curved spatial manifold, not the ordinary coordinate distance $r$ of flat space. Equation (\ref{physrad}) establishes the link between the physical coordinate $r$ in the original $\mathbb{R}^3$ description and the geodesic distance $\rho$ (identified with $R$) on the equivalent constant-curvature space. For $K>0$ ($S^3$ geometry), this mapping is finite-valued: as $r \to \infty$, the geodesic radius approaches its maximum $R_{\text{max}} = \pi/(2\sqrt{\alpha})$. Thus, $R$ measures the actual curved-space “arc length” from the origin, while $r$ is the standard radial coordinate before curvature effects are introduced. This distinction is essential when interpreting uncertainty bounds in curved space, since confinement in $R$ corresponds to a physically different size scale than confinement in $r$.

In the 1D case, corresponding to $S^1$, setting the confinement width to $\Delta x = 2r$ in inequality (\ref{un1d}) and using the definition (\ref{Phi}) of Theorem 1 gives
\begin{equation}
\sigma_p\,\frac{2}{\sqrt{\alpha}}\,\arctan(\sqrt{\alpha}\,r) \ge \pi\hbar.
\end{equation}
With the coordinate transformation (\ref{physrad}) we have $\arctan(\sqrt{\alpha}\,r) = \sqrt{\alpha}\,R$. This substitution cancels the curvature-dependent factors and turns the prefactor into $2R$, the diameter of the geodesic ball on $S^1$. The bound (\ref{un1d}) thus reduces to the simple geodesic representation $\sigma_p\,2R \ge \pi\hbar$, corresponding to the result recently given in \cite{TS22}.\\

Closed-form expressions for the momentum variance bound are only possible in the cases $d=1$ and $d=3$:
For all other dimensions the radial equation involves associated Legendre functions with non-integer degree and non-trivial order, which do not simplify to elementary functions. The eigenvalue condition thus remains transcendental and the corresponding minimal momentum variance must be carried out numerically.

\section{General $d$-Dimensional Operational Inequalities} \label{gen}

In order to apply the approach of the previous section to higher dimensions the question is: If we keep the orthonormality condition $g(X_i, X_j) = \delta_{ij}$ but drop the Killing property, can the variance formula \eqref{VarP} still be maintained in a self-adjoint formulation?
If the Killing property is dropped, the vector fields $X_i$ in general satisfy
\begin{equation}
	\mathrm{div} X_i \neq 0.
\end{equation}
In order to ensure that the associated momentum operators remain self-adjoint with respect to the natural measure $d\mu$, one must employ the De~Witt-corrected definition:
\begin{equation}
	P_i = -i\hbar\left( \nabla_{X_i} + \frac{1}{2} \, \mathrm{div} \, X_i \right).
\end{equation}
With this modification, the operators $P_i$ are formally self-adjoint, and the total momentum variance retains the additive component form (\ref{VarP}) due to the orthonormality of the $X_i$.
However, without the Killing property, the sum $\sum_{i=1}^d P_i^2$ does not in general reduce to $-\hbar^2 \Delta_g$. Instead, one obtains
\begin{equation}\label{deWittVar}
	\sum_{i=1}^d P_i^2 = -\hbar^2 \Delta_g + W[X],
\end{equation}
where $W[X]$ is an additional scalar potential term depending on the frame choice and its divergences  \cite{dewitt52,dewitt57}. This term vanishes identically if and only if each $X_i$ is Killing.

Thus, in $d$ dimensions, the variance formula \eqref{VarP} remains valid in the De~Witt formalism under the sole assumption of orthonormality of the $\{X_i\}$, even without the Killing property. The price to pay is the appearance of an extra potential-like term $W[X]$ in the operator $\sum_{i=1}^d P_i^2$.
Losing the Killing property turns the clean constant-curvature Laplace problem into a variable-coefficient Schrödinger problem with both drift and potential terms, plus a frame-dependence that spoils the universality of the bound. This makes finding the minimal variance much harder: analytically, because symmetry-based separation of variables no longer applies, and conceptually, because the bound may no longer be determined solely by geometry, but also by the specific orthonormal frame chosen.

We now introduce the following De~Witt-symmetrized total momentum variance definition, which—so far as we are aware—has not been explicitly stated in this form, and which underlies the derivations to follow.\\
\\
{\bf Definition.} The \emph{symmetrized total momentum variance} is defined by
 \begin{equation}
 	\sigma_p^2[\psi]
 	= \frac12 \sum_{i,j=1}^d \left\langle (p_i-\langle p_i\rangle)\,g^{ij}\,(p_j-\langle p_j\rangle)
 	+ (p_j-\langle p_j\rangle)\,g^{ij}\,(p_i-\langle p_i\rangle) \right\rangle.
 \end{equation}
Here, the inverse metric $g^{ij}$ acts by multiplication and the subtraction of $\langle p_i\rangle$ ensures that $\sigma_p^2$ is the \emph{covariance} of the momentum components. 
Using the fact that $\langle p_i\rangle$ are scalar numbers commuting with all operators, and employing De~Witt's symmetric ordering identity
\begin{equation}
	\frac12 \left( p_i g^{ij} p_j + p_j g^{ij} p_i \right) = -\hbar^2 \,\Delta_g,
\end{equation}
where $\Delta_g$ is the Laplace--Beltrami operator, we obtain the form
\begin{equation}
	\sigma_p^2[\psi] = -\hbar^2\,\langle \Delta_g \rangle
	- \langle p_i\rangle \,\langle g^{ij}\rangle\,\langle p_j\rangle.
\end{equation}
In contrast to the approach based on (\ref{deWittVar}), this definition renders the momentum variance coordinate-free and compatible with De~Witt's symmetric ordering. The symmetric placement of $g^{ij}$ guarantees that the leading operator is exactly the Laplace--Beltrami operator, with no residual curvature potential. When solving the minimization problem\footnote{$H_0^1(B_R)$ denotes the Sobolev space of square-integrable functions on the domain $B_R$ whose first derivatives are also square-integrable and which vanish on the boundary $\partial B_R$.}
 \begin{equation}
 	\min_{\psi \in H_0^1(B_R),\ \langle\psi,\psi\rangle=1} \ \sigma_p^2[\psi],
 \end{equation}
 with the symmetrized momentum variance the term involving the mean momenta,
 \begin{equation}
 	\langle p_i\rangle\,\langle g^{ij}\rangle\,\langle p_j\rangle,
 \end{equation}
 is non-negative, since $\langle g^{ij}\rangle$ is a positive-definite matrix (being the averaged inverse metric).
 For any trial function $\psi$, subtracting its mean momentum vector $\langle p_i\rangle$ reduces $\sigma_p^2$ or leaves it unchanged. One can always construct a new trial function $\tilde{\psi} = e^{-i\,k_j x^j}\,\psi$,
 or the appropriate geodesic phase factor in curved space, that shifts $\langle p_i\rangle$ to zero without changing the Dirichlet boundary norm. This transformation alters only the mean momentum term and does not affect the kinetic-energy contribution $-\,\hbar^2\langle\Delta_g\rangle$.
 Because the mean-momentum term enters $\sigma_p^2$ with a minus sign and is quadratic in $\langle p_i\rangle$, the minimizing state will always satisfy
 \begin{equation}
 	\langle p_i\rangle_{\mathrm{min}} = 0.
 \end{equation}

Therefore, the minimization problem reduces to finding the ground state of the Laplace-Beltrami operator with Dirichlet boundary conditions on the given region, without the extra mean-momentum term in the variance functional. This ground state can be chosen real and positive, and its momentum variance equals $\hbar^2$ times the first Dirichlet eigenvalue of the Laplacian.

\subsection{Application of the Symmetrized Momentum Variance to the Sphere $S^2_a$}
 
We apply the symmetrized momentum variance definition to the $2$-sphere of radius $a$, denoted $S^2_a$, with Dirichlet boundary conditions on a geodesic ball $B_R \subset S^2_a$.
In spherical coordinates $(\theta,\phi)$, the metric and measure read
\begin{equation}
 	ds^2 = a^2\left(d\theta^2 + \sin^2\theta\,d\phi^2\right), \quad
 	g^{\theta\theta} = \frac{1}{a^2}, \quad g^{\phi\phi} = \frac{1}{a^2\sin^2\theta}, \quad
 	d\mu = a^2\sin\theta\,d\theta\,d\phi.
\end{equation}
 The De~Witt covariant momentum operators in the coordinate basis are
 \begin{equation}
 	p_\theta = -i\hbar\left(\partial_\theta + \frac12\cot\theta\right), \quad
 	p_\phi = -i\hbar\,\partial_\phi.
 \end{equation}
and the Laplace--Beltrami operator on $S^2_a$ is 
 \begin{equation}
 	\Delta_{S^2_a} = \frac{1}{a^2}\left(\partial_\theta^2 + \cot\theta\,\partial_\theta + \frac{1}{\sin^2\theta}\,\partial_\phi^2\right).
 \end{equation}
 The phase-elimination argument shows that for fixed amplitude $|\psi|$ the variance is minimized when the phase is constant, implying $\langle p_i\rangle_{\min} = 0$. Therefore, the minimization problem reduces to the Rayleigh quotient of $-\Delta_{S^2_a}$ on $B_R$:
 \begin{equation}
 	\sigma_{p,\min}^2(R) = \hbar^2 \lambda_1^{(2)}(B_R),
 \end{equation}
 where $\lambda_1^{(2)}(B_R)$ is the first Dirichlet eigenvalue of $-\Delta_{S^2_a}$ on the spherical cap $B_R$.
 The minimizing eigenfunction is azimuthally symmetric, $\psi(\theta,\phi) = \psi(\theta)$. Let $\Theta = R/a$ be the geodesic angular radius. The reduced eigenvalue problem reads
 \begin{equation}
 	\left[ \partial_\theta^2 + \cot\theta\,\partial_\theta + \lambda\,a^2 \right] \psi(\theta) = 0, \quad \psi'(0) = 0, \quad \psi(\Theta) = 0.
 \end{equation}
 With $x = \cos\theta$, this becomes the Legendre equation of order zero. The regular solutions are $P_\nu(\cos\theta)$ with
 \begin{equation}
 	\lambda = \frac{\nu(\nu+1)}{a^2}, \quad P_\nu(\cos\Theta) = 0.
 \end{equation}
 The minimal variance is thus
 \begin{equation}
 	\sigma_{p,\min}^2(R) = \hbar^2\frac{\nu_1(\Theta)\left[\nu_1(\Theta)+1\right]}{a^2},
 \end{equation}
 where $\nu_1(\Theta)$ is the smallest positive solution of $P_\nu(\cos\Theta)=0$. 
As noted above, no closed-form expression for $\lambda_1$ is available in this case. However, a comprehensive asymptotic analysis relevant to our setting is proved in Borisov–Freitas \cite{BF2017} (Thm.\,4.1).

\section{Summary and Outlook}\label{con}

This work reformulates the Extended Uncertainty Principle in explicitly operational, apparatus-centered terms. Instead of expressing ``position uncertainty'' through the statistical width of a wavefunction, we link it to directly tunable features of the experimental apparatus---such as a slit of fixed width in one dimension, or a geodesic ball of given radius in two and three dimensions. The resulting inequalities connect curvature-scale effects, encapsulated by the deformation parameter $\alpha$, to measurable lower bounds for the momentum spread when the preparation size is fixed. In the limit $\alpha \to 0$, these bounds reduce smoothly to the standard Heisenberg uncertainty relation.

The derivations employ self-adjoint momentum operators compatible with the EUP algebra and with Dirichlet boundary conditions matching the physical geometry. In one and three dimensions, closed-form lower bounds are obtained; in two dimensions, the bound is determined by the first Dirichlet eigenvalue of the Laplace--Beltrami operator on a spherical cap, computable numerically with analytically controlled limits.\\
\\
From an experimental perspective, it is crucial to distinguish two regimes:

\begin{enumerate}
	\item \textbf{Laboratory-scale or analogue scenarios with large or effective $\alpha$} (many orders of magnitude above $(H_0/c)^2$):  
	In this case, the bounds derived here can be tested directly with existing or near-term platforms:
	\begin{itemize}
		\item \textbf{1D}: Matter-wave diffraction with tunable slit widths, such as Talbot--Lau or nanograting interferometry using fullerenes or large biomolecules. Here, $\Delta x$ is defined by the physical slit geometry, and $\sigma_p$ is extracted from the diffraction pattern.
		\item \textbf{2D}: Cold-atom traps with controllable circular confinement, probed via release--recapture sequences of Bose--Einstein condensates. The geodesic radius $R$ of the trap plays the role of the apparatus-defined position uncertainty.
		\item \textbf{3D}: Isotropic confinement of ultracold atoms in optical box traps or levitated nanoparticles in optical/magnetic spheres, where $R$ defines the spherical confinement.
	\end{itemize}
	In all these settings, measuring $\sigma_p$ as a function of $\Delta x$ or $R$ provides direct \emph{upper bounds} on $\alpha$. This approach does not assume a cosmological value of $\alpha$ and is sensitive to engineered or effective curvature scales.
	
	\item \textbf{Direct detection of cosmological-scale $\alpha \simeq (H_0/c)^2$}:  
	For $\alpha$ tied to the Hubble parameter, the characteristic length scales required for even a $1\%$ deviation from the standard Heisenberg bound are of order $10^{25}\,\mathrm{m}$ (billions of light-years) for both $\Delta x$ and $R$. Such scales are entirely beyond reach in terrestrial or space-based experiments.  
	
	Space-based interferometry platforms (e.g.\ MAIUS-2, future microgravity missions) offer longer interrogation times and lower environmental noise, but their achievable apparatus sizes are more than $20$ orders of magnitude too small to detect a cosmological $\alpha$ directly. Their realistic role is therefore in constraining \emph{effective} $\alpha$ in analogue systems, not in probing the true cosmological parameter.
\end{enumerate}

The most promising near-term program is to use the operational EUP bounds as precision tools for determining \emph{upper limits on $\alpha$} in controllable laboratory geometries, complemented by analogue models where curvature parameters are deliberately enhanced. This strategy combines immediate experimental feasibility with theoretical readiness for future contexts in which much larger spatial scales—or engineered curvature—can be realised. An empirical violation of these bounds would falsify an entire class of EUP models; confirmation would link $\alpha$ directly to measurable apparatus parameters, opening a route from quantum geometry to precision laboratory physics.

	\bibliographystyle{plain}

\end{document}